# AUDITORY NEURAL RESPONSE INSPIRED SOUND EVENT DETECTION BASED ON SPECTRO-TEMPORAL RECEPTIVE FIELD

*Deokki Min, Hyeonuk Nam, Yong-Hwa Park*

Korea Advanced Institute of Science and Technology, South Korea
{minducky, frednam, yhpark}@kaist.ac.kr

**ABSTRACT**

Sound event detection (SED) is one of tasks to automate function by human auditory system which listens and understands auditory scenes. Therefore, we were inspired to make SED recognize sound events in the way human auditory system does. Spectro-temporal receptive field (STRF), an approach to describe the relationship between perceived sound at ear and transformed neural response in the auditory cortex, is closely related to recognition of sound. In this work, we utilized STRF as a kernel of the first convolutional layer in SED model to extract neural response from input sound to make SED model similar to human auditory system. In addition, we constructed two-branched SED model named as Two Branch STRFNet (TB-STRFNet) composed of STRF branch and baseline branch. While STRF branch extracts sound event information from auditory neural response, baseline branch extracts sound event information directly from the mel spectrogram just as conventional SED models do. TB-STRFNet outperformed the DCASE baseline by 4.3% in terms of threshold-independent macro F1 score, achieving 4$^{th}$ rank in DCASE Challenge 2023 Task 4b. We further improved TB-STRFNet by applying frequency dynamic convolution (FDYConv) which also leveraged domain knowledge on acoustics. As a result, two branch model applied with FDYConv on both branches outperformed the DCASE baseline by 6.2% in terms of the same metric.

***Index Terms***— Sound event detection, STRF, Auditory scene analysis, Human auditory system, auditory neural response

## 1. INTRODUCTION

Sound event detection (SED) is a task for recognition of sound event class and their corresponding time onset and offset [1-4]. SED is closely related to human auditory perception, in that recognizing sound events and their respective time information is essential for the understanding of surrounding acoustic context. Therefore, we were inspired to improve SED by exploiting findings from auditory scene analysis (ASA), a field that aims to translate complex acoustic scene into auditory perception representations within human brain [5]. As sound passes through each part of auditory system, it is transformed into meaningful neural responses by which the auditory cortex can comprehend the perceptual meaning through several steps [6]. The processes include nonlinear amplification, frequency analysis, transformation from vibration into electric signal and higher-order neural computation [7-8]. While aforementioned steps are widely studied and applied to various audio and speech processing works [9-11], the transformation of sound stimulus into auditory cortical neural response is still not entirely comprehended and remains as the subject of ongoing research [12-14].

One approach to simulate the process of transformation from sound stimulus to auditory cortical neural response is to use spectro-temporal receptive field (STRF). STRF is defined as descriptive linear function which predicts primary auditory cortex (A1) cell response for given time-frequency representation of the sound [15, 16]. To estimate STRF, several methods such as spike-triggered average using reverse correlation [17], boosting [18] and machine-learning method such as SVM [19] have been applied to experimental data. Observation on A1 cell response and estimated STRF has revealed that A1 cells have modulation-reactive characteristic that they are easily activated by ripple stimulus which is temporally and spectrally modulated signal [20]. Such spectro-temporal modulation are known to mediate analysis of sound such as speech so that we can obtain the sound intelligibility [21]. In Chi et al [20], STRF is constructed considering such reactive characteristic to dynamic modulation so that STRF captures the spectro-temporal modulation. While some works used the constructed STRF on deep learning applications to extract perceptually important characteristic [22-24], STRF is yet to be applied on SED to the best of our knowledge.

We applied STRFNet proposed by Vuong et al [22], which uses STRF as a convolution kernel in the first convolutional layer of the convolutional neural network (CNN) to imitate the neural response of primary auditory cortex (A1), on SED. However, STRFNet concentrates on extracting modulation property that it is not sufficient to extract various information within sound. To tackle the limitation, we propose two-branch model named as Two Branch STRFNet (TB-STRFNet). While STRF branch extracts the neuroscience-inspired dynamic modulation information using STRF kernel, baseline branch uses conventional convolution to extract the complementary time-frequency information which would not be captured by STRF branch. In addition, we apply frequency dynamic convolution (FDYConv) on TB-STRFNet to further improve the performance. While STRFNet is inspired by auditory neural response to the sound, FDYConv is inspired by the physical nature of time-frequency sound representation. FDYConv was shown to perform the best when applied on on both branches of TB-STRFNet. Joint application of TB-STRFNet and FDYConv significantly improved performance over TB-STRFNet, proving compatibility between two methods as well as the importance of considering domain knowledge.

This paper was supported by the Institute of Civil Military Technology Cooperation funded by the Defense Acquisition Program Administration and Ministry of Trade, Industry and Energy of Korean government under grant No. UM22409RD4.

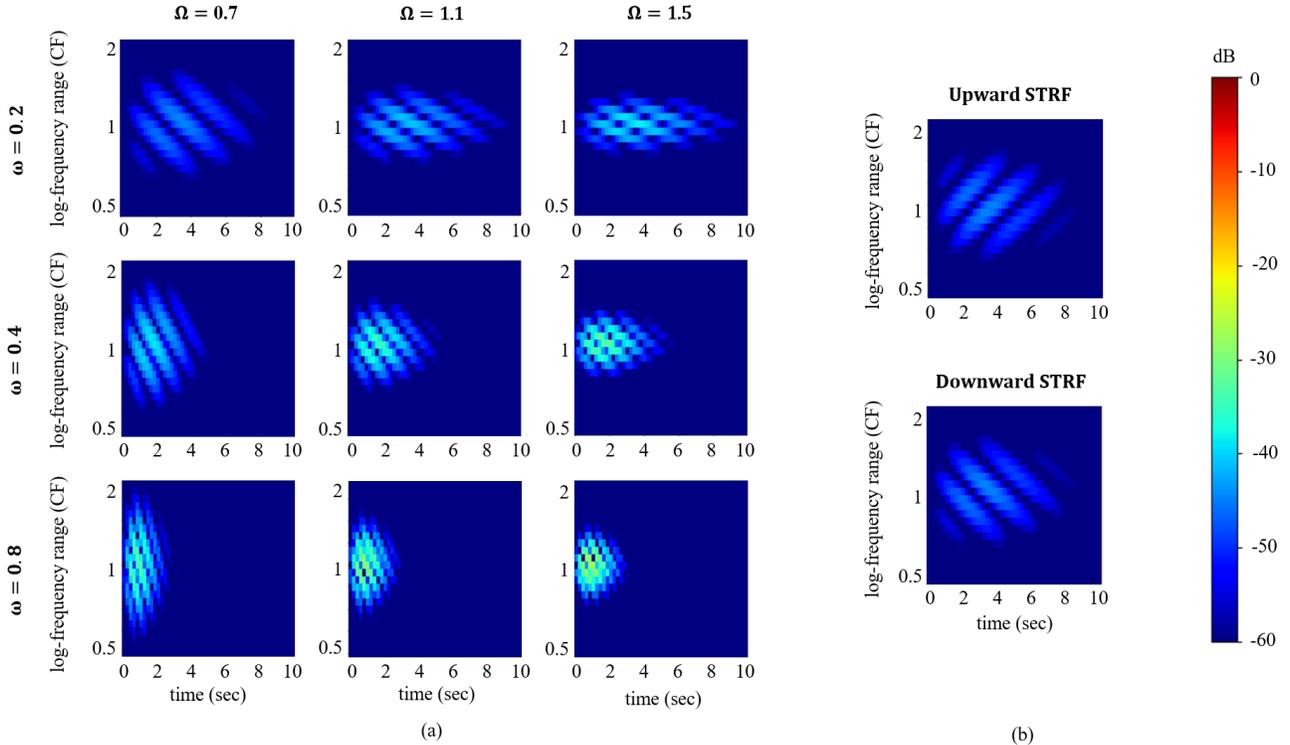

Figure 1. STRF examples with (a) varying scales (Ω) and rates (ω), (b) upward and downward direction.

## 2. PROPOSED METHODS

### 2.1. STRF construction

We adopted STRF construction method by Chi et al [20], where STRF design is abstracted considering particular physiological A1 cell characteristic. Since A1 cell response can be predicted with convolution of sound spectrogram and STRF, STRF should reflect the A1 cell response property. There exists observation on physiological data that A1 cell response is effectively elicited by spectro-temporally modulated ripple [25, 26]. Ripple is defined as a spectro-temporally modulated signal which has temporally varying sinusoidal spectrum along log frequency spacing [27]. Given that A1 cell's modulation-reactive property, STRF needs to be constructed to capture the spectro-temporal modulation. STRF design in Chi et al [20] is a function of both spectral and temporal modulation parameters, which are scale (Ω) and rate (ω) respectively. Scale represents neurons' reaction on range of spectral modulation, while rate represents neurons' reaction on range of temporal modulation. Spectrally and temporally variously tuned neurons could be explained by different combinations of scale and rate values.

Fig. 1 (a) represents constructed STRF examples for varying scale and rate settings. X-axis and y-axis represent time and logarithmic frequency range respectively. STRF is always centered at its center frequency (CF). As 1 CF represents the center frequency, 2 CF and 0.5 CF are double and half of its center frequency respectively. STRF frequency range lies on 2 octaves from 0.5 CF to 2 CF. In fig. 1 (a), scale increases from left column to right column, while rate increases from top to bottom row.

Spectral spacing of ripples is narrower in higher scale while it is wider at lower scale. This illustrates that STRF is narrowly tuned to its center frequency at higher scale while it is broadly tuned at lower scale. Temporal spacing of ripple is narrower in higher rate while it is wider in lower rate. This reflects the characteristics of STRF which is more reactive to impulsive stimulus with higher rate while more reactive to prolonged response time with higher rate. Therefore, the scale variation shows that scale reflects the neural frequency tuning property, while the rate variation shows that rate reflects the neural temporal response property.

STRF has upward and downward direction as shown in fig. 1 (b). While upward direction STRF captures increasing spectral component as time passes, downward STRF captures decreasing spectral component as time passes. Both directions of modulation have to be considered to effectively capture the perceptual meaning of the sound. Note that fig.1 (a) is illustrated as downward STRFs just for consistency.

### 2.2. STRFNet

Constructed STRF has been used as a kernel of the first convolutional layer to tackle several audio-related tasks [22, 23]. We also use this method to verify the effectiveness of STRF on SED. In fig. 2 (a), the architecture of the baseline model used in this work is depicted. Baseline model is composed of six convolution blocks in series followed by two Bi-directional gated recurrent unit (GRU) layers and two fully connected layers. "ConvBlock" in fig. 2 consists of 2D convolutional layer, batch normalization, ReLU activation and 2D maxpool. STRFNet architecture is shown in fig.2 (b), where STRFConv layer is added in front of the baseline model. STRFConv uses 64 different STRFs as con-volutional layer kernels, where 32 STRFs are for upward direction modulation and the other 32 STRFs are for downward direction modulation. Instead of directly training the convolution kernel as in the conventional convolutional layers, STRFConv trains scales and rates corresponding to the channels of the kernel and then produce kernels using trained scales and rates, thus it is lighter than conventional convolutional layers in terms of the number of parameters. In this work, 32 sets of scales and rates are trained in STRFConv to construct 32 upward and downward sets of STRF.

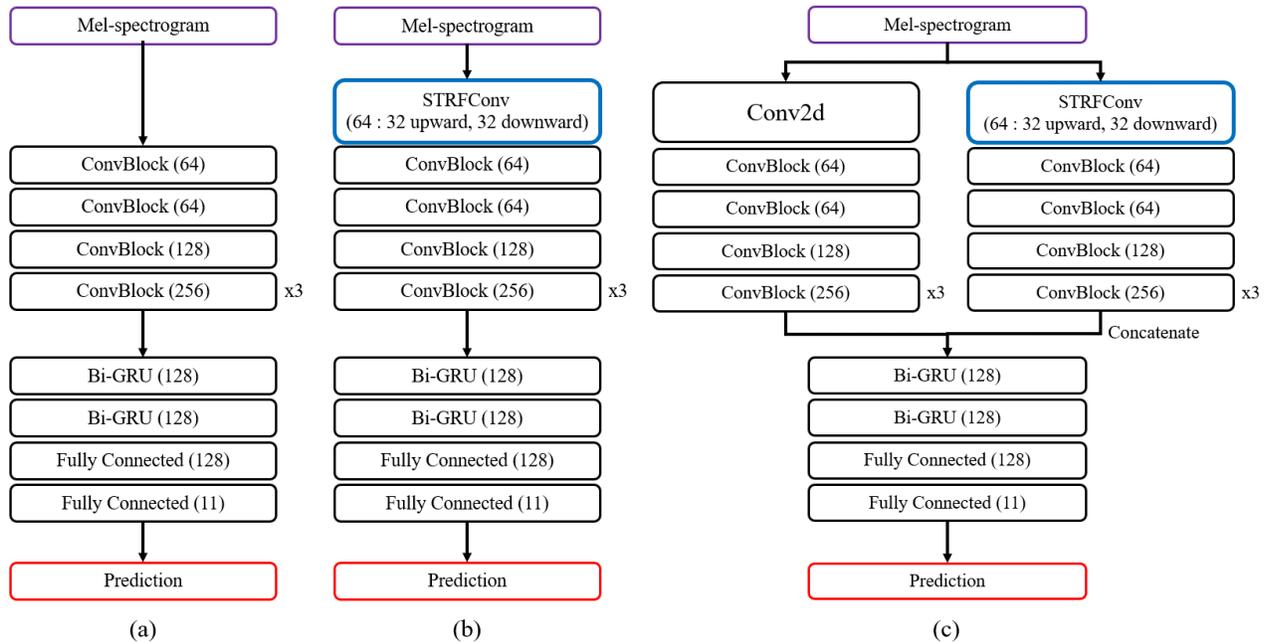

Figure 2: Architectures of (a) baseline model, (b) STRFNet and (c) Two-Branch STRFNet (TB-STRFNet).

### 2.3. TB-STRFNet

STRF kernel only extract spectral and temporal modulation from the sound, while other characteristics of time-frequency patterns within sound would be required for recognition of sound events. In addition, STRF kernel has large kernel size of 50 by 48 and can capture only monotonic modulation form. To compensate such limitations of STRFConv, we propose two-branch model composed of STRF branch and baseline branch, and named as TB-STRFNet. STRF branch is taken from STRFNet while baseline branch is taken from the baseline model with additional 2D convolutional layer before the CNN structure as shown in fig. 2 (c).

Both branches take the identical input which is mel-spectrogram. The STRF branch captures the neuroscience inspired spectro-temporal modulation information from the mel spectrogram. On the other hand, baseline branch consists of a 2D convolutional layer and six convolution blocks. Since STRF branch applies large kernels to extract modulation-related sound information, we expect the baseline branch to extract complementary information using small kernels by focusing on detailed time-frequency patterns. Extracted features from two branches are concatenated to combine separately extracted information.

Concatenated feature map would go through the remaining layers with the same procedure as the baseline model.

### 2.4. STRF with frequency dynamic convolution

To further improve the performance of STRFNet and TB-STRFNet, we experimented on application of frequency dynamic convolution (FDYConv) on those models. FDYConv is proposed to tackle the problem that 2D convolutional layer applies translational equivariance on the frequency dimension while the frequency dimension is shift-variant [9]. This is also related to human auditory system in that it can distinguish frequency-wise translation. Thus FDYConv make sense with the idea of making SED models function similar to human auditory system. To test compatibility of FDYConv with STRFNet, we replaced all convolution layers in STRFNet by FDYConv and named as STRF-FDYNet. In addition, we applied FDYConv on TB-STRFNet to further improve the performance. We replaced convolution layers by FDYConv in baseline branch only, STRF branch only, and both branches of TB-STRFNet and named as TB-STRF-FDYNet1, TB-STRF-FDYNet2, and TB-STRF-FDYNet3, respectively. FDY replaced only convolution layer in convolution blocks for each branch of models, so that first layer of each branch is not FDYConv just as in the original implement of FDYConv [9].

## 3. EXPERIMENTAL DETAILS

### 3.1. Implementation Details

MAESTRO Real dataset is composed of 49 audio clips with duration of 3 to 5 minutes and sampling rate of 44.1 kHz [28]. Mel-spectrogram is used as input feature with 8,820 hop length, 17,640 window length and 64 mel-bin. For training, epoch num-ber is 150, batch size is 32, mean-square error for loss function and Adam optimizer are used. 5 cross-fold validation setup is used for stable overall evaluation.

Table 1: Performance on various SED models.

| Model | Params | F1$_{MO}$(%) |
|---|---|---|
| DCASE baseline [29] | 0.38M | 42.91 |
| baseline | 2.22M | 43.76 |
| TDY-CRNN | 7.01M | 43.57 |
| FDY-CRNN | 7.01M | 44.06 |
| STRFNet | 2.25M | 43.19 |
| STRF-FDYNet | 7.24M | **44.33** |
| TB-baseline | 4.08M | 44.28 |
| TB-STRFNet | 4.08M | 44.75 |
| TB-STRF-FDYNet1 | 9.06M | 44.81 |
| TB-STRF-FDYNet2 | 9.06M | 45.16 |
| TB-STRF-FDYNet3 | 14.05M | **45.55** |

### 3.2. Other SED models for comparison

For comparison, various models are adopted. DCASE baseline model is provided by DCASE Challenge 2023 Task4 subtask B baseline [29]. It has simple model architecture with three CNN layers, one Bi-directional GRU, followed by two fully connected layers. Other than the DCASE baseline model, the models with other methods are based on the baseline model in fig. 2 (a). Temporal dynamic convolution (TDYConv) [11] and frequency dynamic convolution (FDYConv) [9] are dynamic convolution models whose CNN kernel is weighted with time-wise attention and frequency-wise attention respectively. Each temporal and spectral axis-wise attention is extracted from the convolution input. TDY-CRNN and FDY-CRNN are applied in this work to compare the performance with STRF-based models, as they also function similar to human auditory perception and show decent performances. For both TDY-CRNN and FDY-CRNN, dynamic convolutional layer replaced all convolutional layers of the baseline except for the first layer.

### 3.3. Evaluation Metrics

Macro-average F1 score with optimum threshold (F1$_{MO}$) is used for main evaluation metric of DCASE 2023 Task 4 subtask B [30, 31]. By finding the best threshold which is most fit to certain task, the metric can provide more accurate system evaluation and reduce the need of manual threshold optimization. For comparison of model performance, the performance of each model is averaged by 10 sessions, in that 5 cross-validation procedure is performed for one session.

## 4. RESULTS AND DISCUSSION

SER performance of various models discussed in this paper are listed Table 1. For performance of single branch models, TDY-CRNN fails to outperform the baseline model. Since the models apply bi-GRU which considers sequential information between time frames, TDY-CRNN did not improve much compared to the baseline. On the other hand, FDY-CRNN which releases translational equivariance of frequency dimension showed significant improvement. STRFNet which captures the spectro-temporal modulation information, performed worse than the baseline. As we expected, large kernel size of STRF missed the detailed time-frequency information and lead to lower performance. However, joint application of STRFConv and FDYConv has further improved the performance of FDY-CRNN. Considering that STRFConv worsen the performance of baseline while enhance the performance of FDY-CRNN, there exists a synergy between STRFNet and FDYConv. The synergy seems to be due joint application of two methods consistent to principles by human auditory system.

TB-STRFNet which aims to capture detailed time-frequency information while extracting spectro-temporal modulation information, outperformed the baseline and the other single branch models. To verify the effect of increased model size, we constructed TB-baseline, which is consist of two baseline branches. While TB-baseline has almost the same number of parameters with TB-STRFNet, TB-STRFNet outperforms TB-baseline. Since TB-baseline outperforms the baseline, increased model size has affected the model performance. However, considering that application of STRFConv on single branch model has worsened performance, positive effect by STRFConv on TB-STRFNet is apparent. STRF effectively extracts additive information from the mel-spectrogram which is helpful to discriminate the event class and its time onset/offset in two branch architecture. In addition, we interpret that TB-STRFNet outperforms TB-baseline because different role of TB-STRFNet branches efficiently extract the various information from the input sound stimulus. Proposed TB-STRFNet is submitted to DCASE Challenge 2023 Task 4b and achieved 4[th] rank outperforming DCASE baseline by 4.3% [32].

All three models with join application of FDYConv and TB-STRFNet outperformed TB-STRFNet. Since both FDYConv has improved both baseline model and STRFNet, it is effective whether STRFConv is applied or not. In addition, TB-STRF-FDYNet2 applying FDYConv on STRF branch performed better than TB-STRF-FDYNet1 applying FDYConv on the baseline branch. This proves the synergy between STRFConv and FDYConv again, as they perform better when applied together than when applied separately. TB-STRF-FDYNet3 performed the best, improving the baseline by 6.2%. This again proves that methods consistent to human auditory systems are effective on SED and using them together even results in great synergy.

## 5. CONCLUSION

In this work, we applied STRF as a convolutional layer kernel on SED to build SED model functioning closer to human auditory system. While STRFNet performed not as good as the baseline model, TB-STRFNet outperformed the baseline and showed the effect of extracting spectral and temporal modulation information on SED. Furthermore, reflecting frequency-varying perceptual property of auditory system, we applied FDYConv together with STRF. The superior performance of STRF-FDYNet and TB-STRF-FDYNet3 proves that a physiologically consistent deep learning methods enhance SED performance. For future works, we suggest to consider further physiological A1 cell response properties. STRF has dynamic property that STRF is known to be dependent to input sound stimulus [33]. Thus, we may construct the dynamic STRF based model which consider such dynamic STRF property.